\documentclass[
aps,%
12pt,%
final,%
notitlepage,%
oneside,%
onecolumn,%
nobibnotes,%
nofootinbib,
superscriptaddress,%
noshowpacs,%
centertags]%
{revtex4}
\usepackage[cp1251]{inputenc}
\usepackage[T2A]{fontenc}
\usepackage[english,russian]{babel}
\def\lsim{\mathrel{\rlap{
\lower4pt\hbox{\hskip-3pt$\sim$}}
    \raise1pt\hbox{$<$}}}     
\def\gsim{\mathrel{\rlap{
\lower4pt\hbox{\hskip-3pt$\sim$}}
    \raise1pt\hbox{$>$}}}     
\def\scr#1{\mbox{\scriptsize #1}}
\usepackage{graphicx}
\begin{document}
\selectlanguage{english}  
\input{epsf}
\title{Evolution of Baryon-Free Matter \\ Produced
in Relativistic Heavy-Ion Collisions}
\author{\firstname{ V.N.~Russkikh} }
\email[]{Y.Ivanov@gsi.de}
\altaffiliation{Kurchatov Institute, Kurchatov sq.$\!$ 1, Moscow
123182, Russia}
\affiliation{Gesellschaft f\"ur Schwerionenforschung mbH, Planckstr.$\!$ 1,
64291 Darmstadt, Germany  }
\author{\firstname{ Yu.B.~Ivanov} }
\email[]{Y.Ivanov@gsi.de}
\altaffiliation{Kurchatov Institute, Kurchatov sq.$\!$ 1, Moscow
123182, Russia}
\affiliation{Gesellschaft f\"ur Schwerionenforschung mbH, Planckstr.$\!$ 1,
64291 Darmstadt, Germany }
\author{\firstname{ E.G.~Nikonov} }
\email[]{Toneev@gsi.de}
\altaffiliation{Joint Institute for Nuclear Research,
 141980 Dubna, Moscow Region, Russia}
\affiliation{Gesellschaft f\"ur Schwerionenforschung mbH, Planckstr.$\!$ 1,
64291 Darmstadt, Germany  }
\author{\firstname{ W.~N\"orenberg}}
\email[]{W.Nrnbrg@gsi.de}
\affiliation{Gesellschaft f\"ur Schwerionenforschung mbH, Planckstr.$\!$ 1,
64291 Darmstadt, Germany }
\author{\firstname{V.D.~Toneev}}
\email[]{Toneev@gsi.de}
\altaffiliation{Joint Institute for Nuclear Research,
 141980 Dubna, Moscow Region, Russia}
\affiliation{Gesellschaft f\"ur Schwerionenforschung mbH, Planckstr.$\!$ 1,
64291 Darmstadt, Germany  }
\begin{abstract}
A 3-fluid hydrodynamic model is introduced for simulating 
heavy-ion collisions at
incident energies between few and about 200 A$\cdot$GeV. In addition
to the two baryon-rich fluids of 2-fluid models, a new model
incorporates a third,  baryon-free (i.e. with zero net baryonic charge)
fluid which is created in the mid-rapidity region. Its evolution is
delayed due to a formation time $\tau$, during which the baryon-free fluid
neither thermalizes nor interacts with the baryon-rich fluids. After 
formation it thermalizes and starts to interact with the
baryon-rich fluids. It is found that for $\tau=$ 0 the
interaction strongly affects the baryon-free fluid. However, at
reasonable finite formation time, $\tau \simeq$ 1 fm/c, the effect of
this interaction turns out to be substantially reduced although still
noticeable. Baryonic observables are only slightly affected by the
interaction with the baryon-free fluid.
\end{abstract}
\maketitle


\section{Introduction}
Nearly fifty years have passed since the paper
"Relativistic kinetic equation" by Spartak T. Belyaev and Gersh I.
Budker was published~\cite{BB56}. The proposed relativistic
formulation of the distribution function and the kinetic equation
with small-angle scattering
have been included in many textbooks and found numerous
applications in atomic physics and electron-positron plasma.
Recently a generalized relativistic kinetic
equation of this type was implemented for describing the partonic
evolution in very early stages of a heavy-ion collision at
ultra-relativistic RHIC energies~\cite{SST02}. In the present
paper we address more moderate, but nevertheless highly relativistic,
energies, i.e. up to those reached at the CERN SPS. The relativistic
kinetic equation  is used in a peculiar  way, namely to derive a
coupling term for three-fluid hydrodynamic equations.

Two-fluid hydrodynamics with free-streaming radiation of pions was
advanced first in~\cite{MRS88,MRS89}. The initial stage of heavy-ion
collisions definitely is a highly non-equilibrium process.
Within the hydrodynamic approach this
non-equilibrium is simulated by means of a 2-fluid approximation,
which takes care of the finite stopping power of nuclear matter
\cite{IMS85,IS85}, and simultaneously describes the entropy generation
at the initial stage. The radiated pions form a baryon-free matter in the
mid-rapidity region, while two baryon-rich fluids simulate the
propagation of leading particles.  The pions are the most abundant
species of the baryon-free matter which may contain any hadronic and/or
quark-gluon species including baryon-antibaryon pairs.

First applications of the 2-fluid model
\cite{MRS91,MRS91a} to the description of heavy-ion collisions in the
wide range of incident energies, from those of SIS to SPS,
were quite successful. One of the advantages of the hydrodynamic
models is that they directly address the equation of state (EoS) of
 nuclear matter, which is of prime interest for this domain of physics.
In particular, we have shown \cite{INNTS} recently that
the experimental excitation function of the directed flow is well
described by the mixed-phase EoS in contrast to earlier predictions of
the two-phase bag-model EoS. In these 3D
hydrodynamic simulations we describe the whole process of the reaction,
i.e. the evolution from the formation of a hot and dense nuclear
system to its subsequent decay. This is in distinction to numerous 
other simulations, which treat only the expansion stage of a
fireball formed in the course of the reaction, while the initial state of
this dense and hot nuclear system is constructed from
either kinetic simulations or more general albeit
model-dependent assumptions (e.g. see \cite{1-hydro,HS95}).

However, the approximation of free-streaming pions, produced in
the mid rapidity region, was still irritating from the theoretical
point of view, in particular, because the relative momenta of the 
produced pions and the leading baryons are in the range of the $\Delta$
resonance for the incident energies considered. This would imply that the
interaction between the produced pions and the baryon-rich fluids should be
strong. The free-streaming assumption relies on a long formation time 
of  produced pions. Indeed, the proper time for the  formation of the
produced particles   is commonly assumed to be of the order  1 fm/c in
the comoving frame. Since the main part
of the produced pions is quite relativistic at high incident energies,
their formation time should be long enough in the reference frame of
calculation to prevent them from interacting with the baryon-rich fluids.
However, this argument is qualitative rather than
quantitative, and hence requires further verification. The first
attempt to do this was undertaken by the Frankfurt group
\cite{Kat93}, which started to explore an opposite
extreme. They assumed that the produced pions immediately thermalize,
forming a baryon-free fluid (or a ``fireball'' fluid, in terms of
\cite{Kat93}), and interact with the
baryon-rich fluids. No formation time was allowed, and the strength of
the corresponding interaction was guessed rather than microscopically
estimated. This opposite extreme, referred to as a (2+1)-fluid model
and being not quite justified either, yielded results substantially
different from those of the free-streaming approximation. This was one
of the reasons why in subsequent applications the Frankfurt group
neglected the interaction between baryon-free and baryon-rich fluids
while keeping the produced pions thermalized \cite{Brac97}, thus
effectively restoring the free-streaming approximation. However, the
assumed immediate thermalization of the fireball fluid together with
the lack of interaction with baryon-rich fluids still was not a
consistent approximation.

In this paper we would like to return to the problem of
verification of the free-streaming approximation for the produced
pions. To do this, we extend the 2-fluid model of refs.
\cite{MRS88,MRS89,MRS91,MRS91a,INNTS} to a 3-fluid model,
where the created baryon-free fluid (which we call a ``fireball''
fluid, according to the Frankfurt group) is treated on
equal footing with the baryon-rich ones. However, we allow a
certain formation time for the fireball fluid, during which the
constituents of the fluid propagate without interactions.
Furthermore, we estimate the interaction between fireball and
baryon-rich fluids  by means of relativistic kinetic equation and
elementary cross sections.

In this paper we consider  incident energies in the range range from
 few to about 200 A$\cdot$GeV (i.e. from AGS to SPS energies). The
interest to this this energy was recently revived in connection
with the project of the new accelerator facility at GSI SIS200
\cite{SIS200}. The goal of the research program on nucleus-nucleus
collisions at this planned facility is the investigation of
nuclear matter in the region of incident energies
($E_{\scr{lab}}\simeq 10 - 40$ A$\cdot$GeV), in which the highest
baryon densities and highest relative strangeness at moderate
temperatures are expected. Here, the QCD phase diagram is much less
explored, both experimentally and theoretically, as compared to the
higher energy region characterized by higher temperatures, but lower
net  baryon densities, where lattice QCD calculations
\cite{Lattice} and a large body of experimental data from SPS (CERN)
\cite{SPS} and RHIC (BNL) \cite{RHIC} are available by now.

\section{3-Fluid Hydrodynamic Model}

Unlike the one-fluid hydrodynamic model, where local
instantaneous stopping of projectile and target matter is assumed,
a specific feature of the dynamic 3-fluid description is a
finite stopping  power resulting in a counter-streaming regime of leading
baryon-rich matter.  Experimental rapidity distributions in
nucleus--nucleus  collisions support this counter-streaming behavior, which
can be observed for incident energies  between few and 200
A$\cdot$GeV. The basic idea of a 3-fluid approximation to 
heavy-ion collisions \cite{MRS88,MRS89,I87}
is that at each space-time point $x=(t,{\bf x})$
the distribution function of baryon-rich matter, $f_{\scr{br}}(x,p)$, can be
represented as a sum of two distinct contributions
\begin{eqnarray}
\label{t1}
f_{\scr{br}}(x,p)=f_{\scr p}(x,p)+f_{\scr t}(x,p),
\end{eqnarray}
initially associated with constituent nucleons of the projectile (p)
and target (t) nuclei. In addition, newly produced particles,
populating the mid-rapidity region, are associated with a fireball (f)
fluid described by the  distribution function $f_{\scr f}(x,p)$. Note
that both the baryon-rich and fireball fluids may consist of any type
of hadrons  and/or partons (quarks and gluons), rather then only
nucleons and pions. However, 
here and below we suppress the species label at the distribution
functions for the sake of transparency of the equations.

With the above-introduced distribution
functions $f_\alpha$ ($\alpha=$p, t, f), the coupled set of relativistic
Boltzmann equations looks as follows:
\begin{eqnarray}
p_\mu\partial^\mu_x f_{\scr p} (x,p) &=& C_{\scr p} (f_{\scr p},f_{\scr t})+C_{\scr p} (f_{\scr p},f_{\scr f}),
   \label{t2}
\\
p_\mu\partial^\mu_x f_{\scr t} (x,p) &=& C_{\scr t} (f_{\scr p},f_{\scr t})+C_{\scr t} (f_{\scr t},f_{\scr f}),
   \label{t3}
\\
p_\mu\partial^\mu_x f_{\scr f} (x,p) &=& C_{\scr f} (f_{\scr p},f_{\scr t})
+C_{\scr f} (f_{\scr p},f_{\scr f})+C_{\scr f} (f_{\scr t},f_{\scr f}),
   \label{t4}
\end{eqnarray}
where $C_\alpha$ denote collision terms between the constituents of
the three fluids. We have omitted intra-fluid collision terms, like
$C_{\scr p} (f_{\scr p},f_{\scr p})$,
since below they will be canceled any way. The displayed inter-fluid
collision terms have a clear physical meaning:
$C_{\scr p/t} (f_{\scr p},f_{\scr t})$,
$C_{\scr p/t} (f_{\scr p/t},f_{\scr f})$, and
$C_{\scr f} (f_{\scr p/t},f_{\scr f})$ give rise to
friction between p-, t- and f-fluids, and
$C_{\scr f} (f_{\scr p},f_{\scr t})$
takes care of particle production in the mid-rapidity region.
Note that up to now we have done no approximation, except for
hiding intra-fluid collision terms.

Let us proceed to approximations which justify the term ``fluids'' 
having been used already. We assume that constituents
within each fluid are  locally equilibrated, both thermodynamically
and chemically.  In particular, this implies that the 
intra-fluid collision terms are indeed zero.
This assumption relies on the fact that  intra-fluid
collisions are much more efficient in driving a system to equilibrium
than  inter-fluid interactions. As applied to the fireball fluid,
this assumption requires some additional comments, related to the
concept of a finite formation time. During the formation proper time
$\tau$ after production, the fireball fluid propagates  freely,
interacting neither  with itself nor with the baryon-rich
fluids. . After this time period, the fireball matter locally 
thermalizes and  starts to interact with both itself and
the baryon-rich fluids. Being heated up,
 these three fluids may contain not only hadronic and but also
 deconfined  quark-gluon species, depending on the EoS used.

The above assumption suggests that interaction between different fluids
should be treated dynamically. To obtain the required dynamic equations,
we first integrate the kinetic Eqs.~(\ref{t2})--(\ref{t4}) over momentum
and sum over particle species with weight of baryon charge. This way
we arrive to equations of the baryon charge conservation
   \begin{eqnarray}
   \label{eq8}
   \partial_{\mu} J_{\alpha}^{\mu} (x) &=& 0,
   \end{eqnarray}
for $\alpha=$p and t, where
$J_{\alpha}^{\mu}=n_{\alpha}u_{\alpha}^{\mu}$ is the baryon current
defined in terms of baryon density $n_{\alpha}$ and
 hydrodynamic 4-velocity $u_{\alpha}^{\mu}$ normalized as
$u_{\alpha\mu}u_{\alpha}^{\mu}=1$. Eq.~(\ref{eq8}) implies that there
is no baryon-charge exchange between p- and t-fluids, as well as that
the baryon current of the fireball fluid is identically zero,
$J_{\scr f}^{\mu}=0$. Integrating kinetic
Eqs.~(\ref{t2})--(\ref{t4}) over momentum with weight of 4-momentum
$p^\nu$ and  summing over all particle species, we arrive at equations
of the energy--momentum exchange for energy--momentum tensors
$T^{\mu\nu}_\alpha$ of the fluids
   \begin{eqnarray}
   \partial_{\mu} T^{\mu\nu}_{\scr p} (x) &=&
-F_{\scr p}^\nu (x) + F_{\scr{fp}}^\nu (x),
   \label{eq8p}
\\
   \partial_{\mu} T^{\mu\nu}_{\scr t} (x) &=&
-F_{\scr t}^\nu (x) + F_{\scr{ft}}^\nu (x),
   \label{eq8t}
\\
   \partial_{\mu} T^{\mu\nu}_{\scr f} (x) &=&
F_{\scr p}^\nu (x) + F_{\scr t}^\nu (x)
- F_{\scr{fp}}^\nu (x) - F_{\scr{ft}}^\nu (x),
   \label{eq8f}
   \end{eqnarray}
where the $F^\nu$ are friction forces originating from inter-fluid
collision terms in the kinetic Eqs.~(\ref{t2})--(\ref{t4}).
$F_{\scr p}^\nu$ and $F_{\scr t}^\nu$ in
Eqs.~(\ref{eq8p})--(\ref{eq8t}) describe energy--momentum loss of
baryon-rich fluids due to their mutual friction. A part of this loss
$|F_{\scr p}^\nu - F_{\scr t}^\nu|$ is transformed into thermal
excitation of these fluids, while another part
$(F_{\scr p}^\nu + F_{\scr t}^\nu)$ gives rise to particle production
into the fireball fluid (see Eq.~(\ref{eq8f})). $F_{\scr{fp}}^\nu$
and $F_{\scr{ft}}^\nu$ are associated with friction of the fireball
fluid with the p- and t-fluids, respectively.
Note that Eqs.~(\ref{eq8p})--(\ref{eq8f})  satisfy
the  total energy--momentum conservation
\begin{eqnarray}
\partial_\mu (T^{\mu\nu}_{\scr p} +
T^{\mu\nu}_{\scr t} + T^{\mu\nu}_{\scr f}) = 0.
\label{eq10}
\end{eqnarray}

As described above, the energy--momentum tensors of the baryon-rich
fluids ($\alpha=$p and t) take the conventional hydrodynamic form
   \begin{eqnarray}
T^{\mu\nu}_\alpha=
(\varepsilon_\alpha + P_\alpha) \ u_{\alpha}^{\mu} \ u_{\alpha}^{\nu}
-g^{\mu\nu} P_\alpha
   \label{eq11}
   \end{eqnarray}
in terms of the proper energy density, $\varepsilon_\alpha$, and
pressure, $P_\alpha$. For the fireball, however, only
the thermalized part of the energy--momentum tensor is described
by this hydrodynamic form
   \begin{eqnarray}
T^{\scr{(eq)}\mu\nu}_{\scr f}=
(\varepsilon_{\scr f} + P_{\scr f}) \ u_{\scr f}^{\mu} \ u_{\scr f}^{\nu}
-g^{\mu\nu} P_{\scr f}.
   \label{eq12}
   \end{eqnarray}
Its evolution is defined by a Euler equation with a retarded source term
   \begin{eqnarray}
   \partial_{\mu} T^{\scr{(eq)}\mu\nu}_{\scr f} (x) =
\int d^4 x' \delta^4 \left(\vphantom{I^I_I}
x - x' - U_F (x')\tau\right)
 \left[F_{\scr p}^\nu (x') + F_{\scr t}^\nu (x')\right]
- F_{\scr{fp}}^\nu (x) - F_{\scr{ft}}^\nu (x),
   \label{eq13}
   \end{eqnarray}
where $\tau$ is the formation time, and
   \begin{eqnarray}
   \label{eq14}
U^\nu_F (x')=
\frac{F_{\scr p}^\nu(x')+F_{\scr t}^\nu(x')}%
{|F_{\scr p}(x')+F_{\scr t}(x')|}
   \end{eqnarray}
is a free-streaming 4-velocity of the produced fireball matter. In fact,
this is the velocity at the moment of production of the fireball matter.
According to Eq.~(\ref{eq13}), the energy and momentum of this matter
appear as a source in the Euler equation only later, at the time
$U_F^0\tau$ after production, and in different space point
${\bf x}' - {\bf U}_F (x')\tau$, as compared to the production point
${\bf x}'$. From the first glance, one can immediately simplify the r.h.s.
of Eq. (\ref{eq13}) by performing integration with the $\delta$-function.
However, this integration is not that straightforward, since the expression
under the $\delta$-function, $x - x' - U_F (x')\tau=0$, may have more than
one solution with respect to $x'$. The latter would mean that the matter
produced in several different space-time points $x'$ simultaneously
thermalizes in the same space-time point $x$. This is possible due to
the nonlinearity of the hydrodynamic equations.

The residual part of $T^{\mu\nu}_{\scr f}$ (the free-streaming one)
is defined as
   \begin{eqnarray}
T^{\scr{(fs)}\mu\nu}_{\scr f}=
T^{\mu\nu}_{\scr f}-T^{\scr{(eq)}\mu\nu}_{\scr f}.
   \label{eq15}
   \end{eqnarray}
The equation for $T^{\scr{(eq)}\mu\nu}_{\scr f}$ can be easily
obtained by taking the difference between Eqs.~(\ref{eq8f}) and
(\ref{eq13}). If all  the fireball matter turns out to be formed before
freeze-out, than this equation is not needed. Thus, the 3-fluid model
introduced here contains both the original  2-fluid model with
pion radiation \cite{MRS88,MRS89,MRS91,MRS91a,INNTS} and the
(2+1)-fluid model \cite{Kat93,Brac97}
as limiting cases for $\tau \rightarrow \infty$ and $\tau=0$,
respectively.

The nucleon--nucleon cross sections at high energies are strongly
forward--backward peaked. In this case the Boltzmann
collision term  can be essentially simplified,
since the involved 4-momentum transfer is small. The
small-angle-scattering expansion of the
collision integral results in the relativistic Fokker--Planck equation,
as first derived by Belyaev and Budker \cite{BB56}. Precisely this equation
was used in \cite{IMS85} to estimate the friction forces,
$F_{\scr p}^\nu$ and $F_{\scr t}^\nu$, proceeding
from only $NN$ elastic scattering. Later these friction forces were
calculated \cite{Sat90} based on (both elastic and inelastic)
experimental inclusive $NN$ cross sections
 \begin{eqnarray}
 F_{\alpha}^\nu=\rho_{\scr p} \rho_{\scr t}
\left[\left(u_{\alpha}^{\nu}-u_{\bar{\alpha}}^{\nu}\right)D_P+
\left(u_{\scr p}^{\nu}+u_{\scr t}^{\nu}\right)D_E\right],
\label{eq16}
\end{eqnarray}
$\alpha=$p and t, $\bar{\mbox{p}}=$t and  $\bar{\mbox{t}}=$p. Here,
$\rho_\alpha$ denotes the scalar densities of the p- and t-fluids,
 \begin{eqnarray}
D_{P/E} = m_N \ V_{\scr{rel}}^{\scr{pt}} \ 
\sigma_{P/E} (s_{\scr{pt}}),
\label{eq17}
\end{eqnarray}
where $m_N$ is the nucleon mass, $s_{\scr{pt}}=m_N^2 \left(u_{\scr
p}^{\nu}+u_{\scr t}^{\nu}\right)^2$ is the mean invariant energy
squared of two colliding nucleons from the p- and t-fluids,
$V_{\scr{rel}}^{\scr{pt}}=
[s_{\scr{pt}}(s_{\scr{pt}}-4m_N^2)]^{1/2}/2m_N^2$ is the mean
relative velocity of the p- and t-fluids, and
$\sigma_{P/E}(s_{\scr{pt}})$ are determined in terms of
nucleon-nucleon cross sections integrated with certain weights
(see \cite{MRS88,MRS89,MRS91,MRS91a,Sat90} for details). It was
found in \cite{Sat90} that a part of these friction terms, which
is related to the transport cross-section, may be well
parameterized by  an effective deceleration length $\lambda_{\rm
eff}$ with a constant value $\lambda_{\rm eff}\approx 5 $ fm.
However, there are reasons to consider $\lambda_{\rm eff}$ as a
phenomenological parameter, as it was pointed out in \cite{MRS91}.
Indeed, as it is seen from Eq.~(\ref{eq17}), this friction is
estimated only in terms of nucleon-nucleon cross sections while
the excited matter of baryon-rich fluids certainly consists of
great number of hadrons and/or deconfined quarks and
gluons. Furthermore, these quantities may be modified by in-medium
effects. In this respect, $D_{P/E}$ should be understood as quantities
that give a scale of this interaction.

Eqs.~(\ref{eq8})--(\ref{eq8t}) and (\ref{eq13}), supplemented by
a certain EoS and expressions for friction forces $F^\nu$, form a full
set of equations of the relativistic 3-fluid hydrodynamic model. To make
this set closed, we still need to define the friction of the fireball
fluid with the p- and t-fluids, $F_{\scr{fp}}^\nu$
and $F_{\scr{ft}}^\nu$ in terms of hydrodynamic quantities and some
cross sections.

\section{Interaction between Fireball and Baryon-Rich Fluids}

Our aim here is to estimate the scale of the friction force between the
fireball and baryon-rich fluids, similar to that done before
for baryon-rich fluids \cite{Sat90}. To this end, we consider a
simplified system, where all baryon-rich fluids consist only of
nucleons, as the most  the abundant component of these fluids, and
the fireball fluid contains only  pions.

For incident energies from 10 (AGS) to 200 A$\cdot$GeV (SPS),
the relative nucleon-pion energies are in the resonance range
dominated by the $\Delta$-resonance. To estimate this relative
energy we consider a produced pion, being at rest in the center of
mass of the colliding nuclei,$q_\pi=\{m_\pi,0,0,0\}_{cm}$. Baryon-rich
fluids decelerate each other during their interpenetration. This means
that the nucleon momentum should be smaller than the incident momentum,
$|p_N| < |\{m_N\gamma_{cm},{\bf p}_{cm}\}|$, where $\gamma_{cm}$
is the gamma factor of the incident nucleon in the cm frame.
Calculating the invariant relative energy squared $s=(p+q)^2$ at
$E_{\scr{lab}}=$ 158 A$\cdot$GeV, we obtain $s^{1/2}<$ 1.8 GeV.
This range of $s$ precisely covers the resonance region, 1.1 GeV
$<s^{1/2}<$ 1.8 GeV \cite{PPVW93}. At $E_{\scr{lab}}=$ 10
A$\cdot$GeV we arrive at $s^{1/2}<$ 1.3 GeV, which is also within
the resonance region. At even lower incident energies the strength
of the fireball fluid becomes so insignificant, as compared with
thermal mesons in the p- and t-fluids, that the way of treatment
of its interaction with the baryon-rich fluids does not essentially
affect  the observables. For the same reason we do not apply any 
special prescription for the unification of the fireball fluid with
the baryon-rich fluids, since this may happen only at relatively low
incident energies $E_{\scr{lab}} <$ 10 A$\cdot$GeV.

The resonance-dominated interaction implies that the
essential process is absorption of a fireball pion by a p- or
t-fluid nucleon with formation of an $R$-resonance (most
probably $\Delta$). This produced $R$-resonance still belongs to the
original p- or t-fluid, since its recoil due to absorption of a light
pion is small. Subsequently this $R$-resonance 
decays into a nucleon and a pion already belonging to the original p- or
t-fluid. Symbolically, this mechanism can be expressed as
$$ N^\alpha + \pi^{\scr f} \to
R^\alpha \to  N^\alpha + \pi^\alpha .$$ As a consequence,  only the
loss term contributes to the kinetic equation for the fireball fluid.

Proceeding from the above consideration, we write down
the collision term between fireball-fluid pions and  $\alpha$-fluid
nucleons ($\alpha=$p or t) as follows
\begin{eqnarray}
\label{eq18}
C_{\scr f} (f_\alpha,f_{\scr f}) &=& -
\int \frac{d^3 q}{q_0} \  
W^{N\pi\to R}(s) \ f_{\scr f}^{\scr{(eq)}} (x,p) \ f_\alpha (x,q),
\end{eqnarray}
where $s=(p+q)^2$,
$$W^{N\pi\to R}(s) = (1/2)
[(s-m_N^2-m_\pi^2)^2-4m_N^2m_\pi^2]^{1/2} \ 
\sigma_{\scr{tot}}^{N\pi\to R}(s)$$
is the rate to produce a baryon $R$-resonance, and
$\sigma_{\scr{tot}}^{N\pi\to R}(s)$ is the parameterization of experimental
pion--nucleon cross-sections \cite{PPVW93}. Here, only the distribution
function of formed (and hence thermalized) fireball pions,
$f_{\scr f}^{\scr{(eq)}}$, enters the collision term, since the
non-formed particles did not participate in the interaction by assumption.

Integrating $C_{\scr f} (f_\alpha,f_{\scr f})$ 
 weighted with the 4-momentum $p^\nu$ over momentum, we arrive at
\begin{eqnarray}
\hspace*{-7mm}
F_{\scr{f}\alpha}^\nu (x)
&=&
\int \frac{d^3 q}{q_0} \frac{d^3 p}{p_0} p^\nu  W^{N\pi\to R}(s) \
f_{\scr f}^{\scr{(eq)}} (x,p)  \ f_\alpha (x,q)
\cr
&\simeq&
\frac{W^{N\pi\to R}(s_{{\scr f}\alpha})}{m_\pi u_{\scr f}^0}
\left(\int \frac{d^3 q}{q_0} f_\alpha (x,q)\right)
\left(\int \frac{d^3 p}{p_0} p^0 p^\nu f_{\scr f}^{\scr{(eq)}} (x,p)\right)
=
 D_{\scr{f}\alpha}\frac{T^{\scr{(eq)}0\nu}_{\scr f}}{u_{\scr
 f}^0}\rho_{\alpha}~, 
\label{eq19}
\end{eqnarray}
where we substituted $p^0$ and $s$ by their mean values,
$<p^0>=m_\pi u_{\scr f}^0$ and
$s_{{\scr f}\alpha} = (m_\pi u_{\scr f}+m_N u_{\alpha})^2$,
and introduced the transport coefficients
\begin{eqnarray}
D_{\scr{f}\alpha} = W^{N\pi\to R}(s_{{\scr f}\alpha})/(m_N m_\pi)
=V_{\scr{rel}}^{{\scr f}\alpha} \ 
\sigma_{\scr{tot}}^{N\pi\to R}(s_{{\scr f}\alpha}).
\end{eqnarray}
Here, $V_{\scr{rel}}^{{\scr f}\alpha}=[(s_{{\scr
f}\alpha}-m_N^2-m_\pi^2)^2 -4m_N^2m_\pi^2]^{1/2}/(2m_N m_\pi)$ denotes
the mean invariant relative velocity between the fireball and 
the $\alpha$-fluids. Thus, we have expressed the friction
$F_{\scr{f}\alpha}^\nu$ in terms of the fireball-fluid
energy-momentum density $T^{0\nu}_{\scr f}$, the scalar density
$\rho_{\alpha}$ of the $\alpha$ fluid, and a transport coefficient
$D_{\scr{f}\alpha}$. Note that this friction is zero until the
fireball pions are formed, since $T^{\scr{(eq)}0\nu}_{\scr f}=0$
during the formation time $\tau$.

In fact, the above treatment is an estimate of the friction
terms rather than their strict derivation. This peculiar way of
evaluation is motivated by the form of the final result
(\ref{eq19}). An advantage of this form is that $m_\pi$ and any
other mass do not appear explicitly, and hence allows a natural
extension  to any content of the fluid,
including deconfined quarks and gluons, assuming that
$D_{\scr{f}\alpha}$ represents just a scale of the transport
coefficient.

\section{Simulations of Nucleus--Nucleus Collisions}

The relativistic 3D code for the above described 3-fluid model was
constructed by means of modifying the existing 2-fluid 3D code of
refs. \cite{MRS88,MRS89,MRS91,MRS91a,INNTS}. In actual
calculations we used the mixed-phase EoS developed in
\cite{NST98,TNS98,NTS99}. This phenomenological EoS takes into account
a possible deconfinement phase transition of nuclear matter. The
underlying  assumption of this EoS is that  unbound quarks and
gluons may coexist with hadrons in the nuclear environment. In accordance
with lattice QCD data, the statistical mixed-phase model describes
the first-order deconfinement phase transition for pure gluon matter
and crossover for that with quarks  \cite{NST98,TNS98,NTS99}.

We performed simulations of nucleus--nucleus collisions Pb+Pb at
$E_{\scr{lab}}=$ 158 A$\cdot$GeV and Au+Au at $E_{\scr{lab}}=$
10.5 A$\cdot$GeV. General dynamics of heavy-ion collisions is
illustrated  in Fig. 1 by  the energy-density evolution of the
baryon-rich fluids 
($\varepsilon_{\scr{b}}=\varepsilon_{\scr{p}}+\varepsilon_{\scr{t}}$,
in the cm frame of colliding nuclei) in the reaction plane of the
 Pb+Pb collision. Different stages of
interaction at relativistic energies are clearly seen in this
example: Two Lorentz-contracted nuclei (note the different scales
along the $x$- and $z$-axes in Fig.1) start to interpenetrate through each
other, reach 
a maximal energy density by the time $ \sim 1.1 $ fm/c and
then expand  predominantly in longitudinal direction forming a
"sausage-like" freeze-out system. At this and lower incident energies
 the baryon-rich dynamics is not really disturbed by the fireball
fluid and hence the cases $\tau=$ 0 and 1
fm/c turned to be indistinguishable in terms of $\varepsilon_{\scr{b}}$.

In Fig. 2 the dynamic evolution of the fireball energy
density ($\varepsilon_{\scr{f}}$, in the cm frame of the colliding
nuclei) in the reaction plane of the Pb+Pb
collision at impact parameter $b=$ 2 fm is shown for two values of
the  formation time, $\tau=$ 0 fm/c (the  left column of panels)
and 1 fm/c (the right column of panels). It starts to form near the
time moment, when the maximal energy density $\varepsilon_{\scr{b}}$ is
reached. The $f$-fluid 
evolution indeed looks like that for an expanding fireball, its
density essentially depends on the formation time.

\begin{figure}
\begin{center}
\includegraphics[height=17cm,clip]{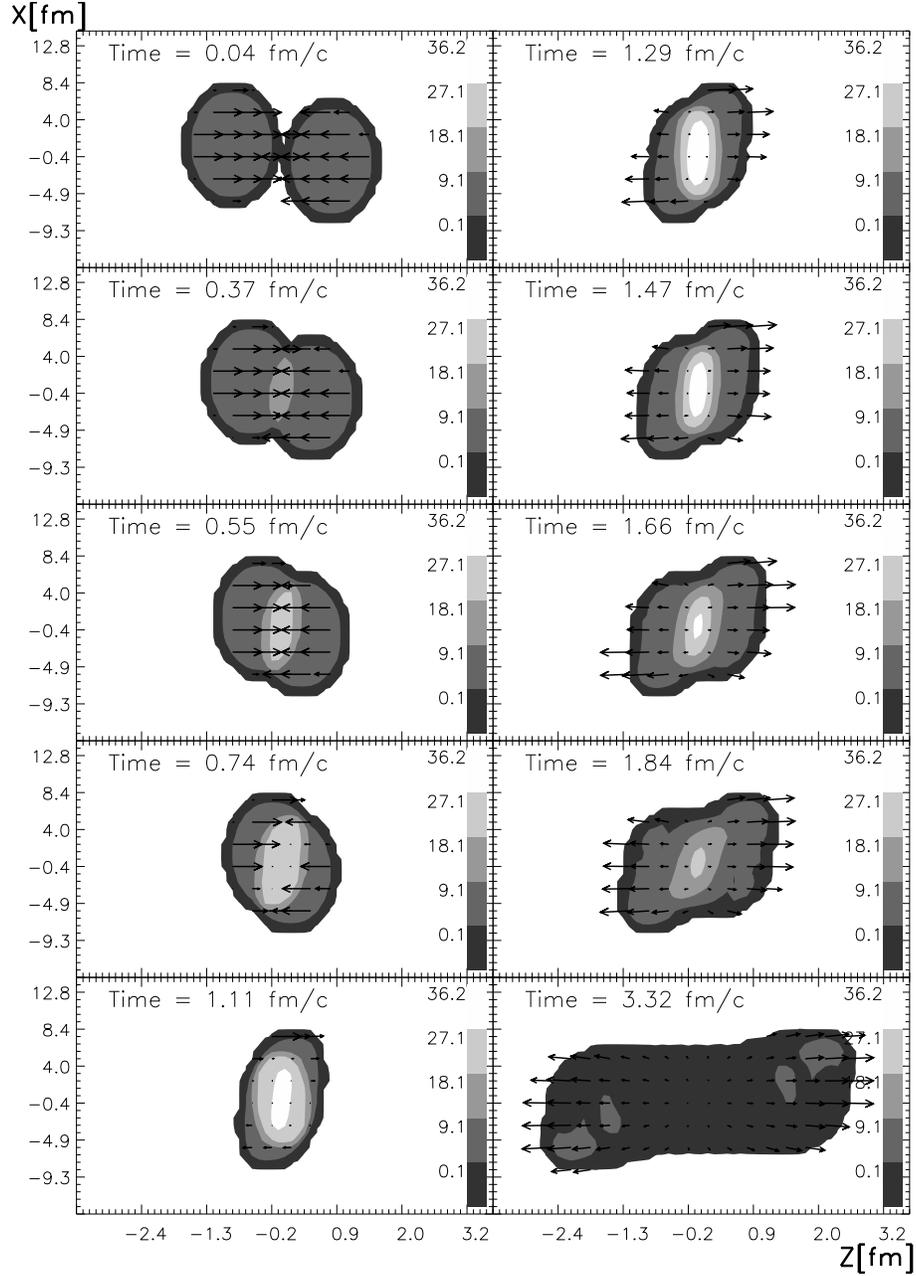}
\caption[C1]{Time evolution of the  energy density, 
$\varepsilon_{\scr{b}}=\varepsilon_{\scr{p}}+\varepsilon_{\scr{t}}$,
 for  the baryon-rich fluids in the 
reaction plane ($xz$ plane) for the Pb+Pb collision
($E_{\scr{lab}}=$ 158 A$\cdot$GeV) at impact parameter $b=$ 2 fm.
Shades of gray represent different levels of
$\varepsilon_{\scr{b}}$ as indicated at the right side of each
panel. Numbers at this palette show the $\varepsilon_{\scr{b}}$
values (in GeV/fm$^3$) at which the shades change. Arrows indicate
the hydrodynamic velocities of the fluids. } \label{fig1}
\end{center}
\end{figure}
\begin{figure}
\begin{center}
\includegraphics[height=17cm,clip]{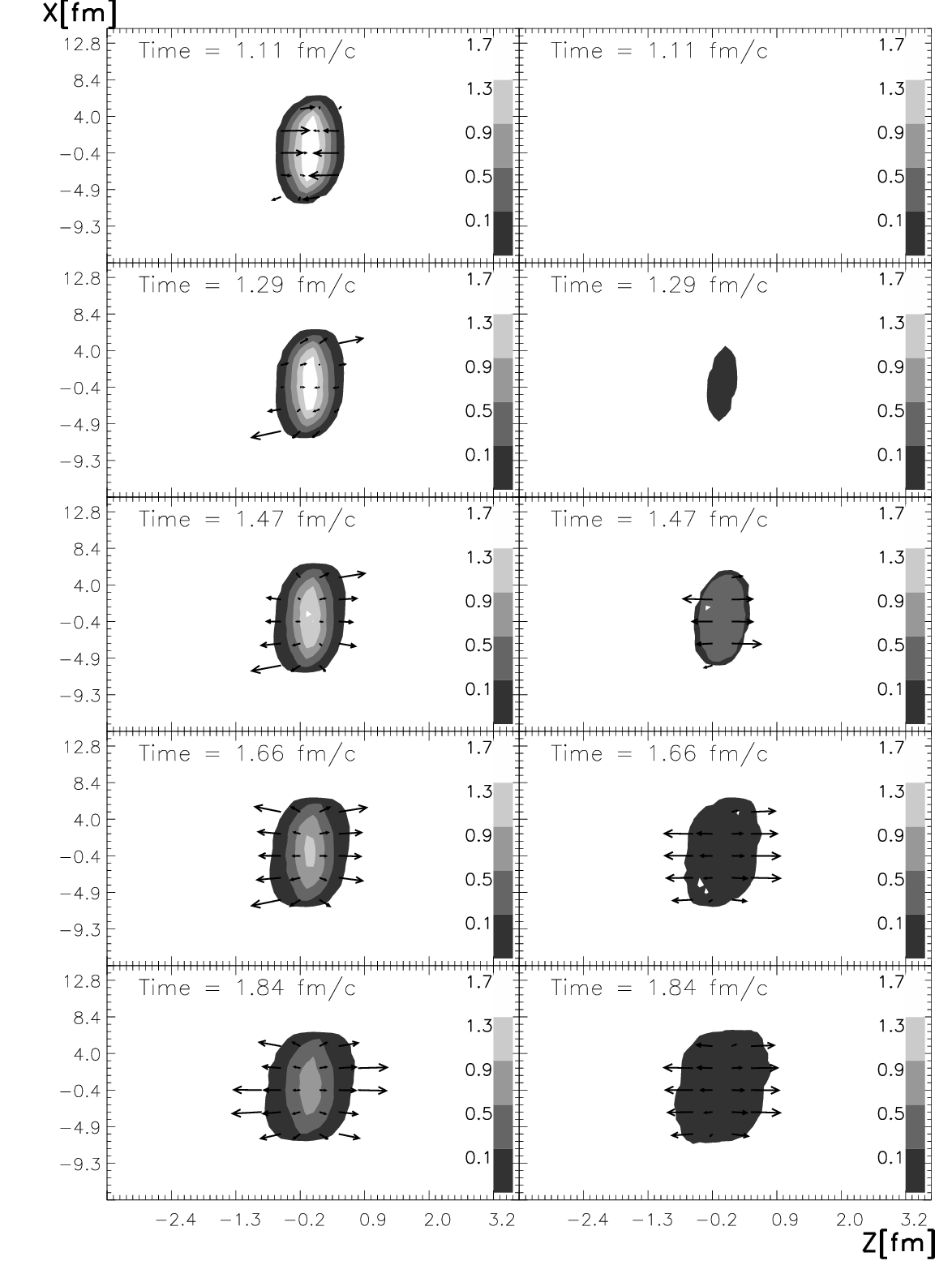}
\caption[C2]{The same as in Fig. 1, but for the fireball-energy
density ($\varepsilon_{\scr{f}}$ in the cm frame of the colliding
nuclei) for two formation times, $\tau=$ 0 fm/c (the left column
of panels) and 1 fm/c (the right column of panels). } \label{fig2}
\end{center}
\end{figure}

To quantitively reveal the role of the interaction between
fireball and baryon-rich fluids, we followed the evolution of the
total energy released into the fireball fluid
\begin{eqnarray}
   \label{eq20}
E_{\scr f}^{\scr{(released)}} (t) = \int_0^t dt' \int d^3 x'
\left(F_{\scr p}^0 (x') + F_{\scr t}^0 (x') \right)~,
\end{eqnarray}
cf. Eq.~(\ref{eq8f}), and the total energy kept in the fireball
fluid (both thermalized and nonthermalized) after interaction,
 \begin{eqnarray}
    \label{eq21}
E_{\scr f}^{\scr{(tot)}} (t)
&=&\int d^3 x \ T^{00}_{\scr f} (t,{\bf x})
\cr
&=& \int_0^t dt' \int d^3 x' \ 
\left(F_{\scr p}^0 (x') + F_{\scr t}^0 (x')
- F_{\scr{fp}}^0 (x') - F_{\scr{ft}}^0 (x')\right),
\end{eqnarray}
cf. Eq.~(\ref{eq8f}), in the cm frame of two colliding nuclei.
Results of the calculation are pre\-sen\-ted in Fig. 3. To provide
a common scale, the $E_{\scr f}^{\scr{(released)}}$ quantity
calculated with the formation time $\tau=$ 100 fm/c is presented
in all the panels. The $\tau=$ 100 fm/c case practically implies
absence of interaction between the fireball and baryon-rich fluids
and the equality $E_{\scr f}^{\scr{(tot)}}=E_{\scr
f}^{\scr{(released)}}$, because $T^{\scr{(eq)}00}_{\scr f}=0$ and
hence $F_{\scr{fp}}^0= F_{\scr{ft}}^0=0$, cf. Eq.~(\ref{eq19}).

\begin{figure}[h]
\begin{center}
\includegraphics[height=12cm,clip]{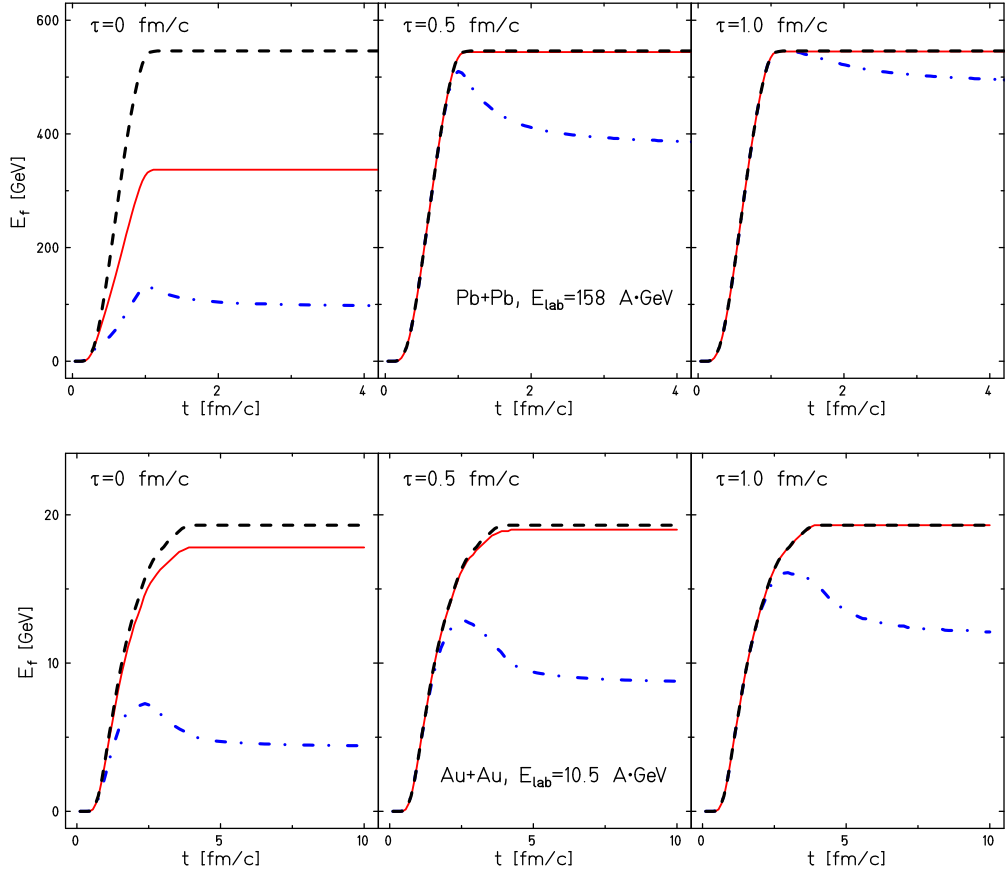}
\caption[C3]{Time evolution of the total energy released into the
fireball fluid, $E_{\scr f}^{\scr{(released)}}$ (solid lines), and
the total energy kept in the fireball fluid after interaction,
$E_{\scr f}^{\scr{(tot)}}$ (dashed-dotted lines).  Two
nucleus--nucleus collisions Pb+Pb at $E_{\scr{lab}}=$ 158
A$\cdot$GeV (three upper panels) and Au+Au at $E_{\scr{lab}}=$
10.5 (three bottom panels) A$\cdot$GeV, both at zero impact
parameter, were calculated with different formation times $\tau$
indicated in the panels. The upper dashed curve in all 6 panels
represent $E_{\scr f}^{\scr{(released)}}$ calculated with the
formation time $\tau=$ 100 fm/c.} \label{fig3}
\end{center}
\end{figure}

First, we see that the energy release into in the fireball fluid
occurs only during a short time of interpenetrating of colliding
nuclei. As it has been expected, at zero formation time $\tau=$ 0
the interaction with the baryon-rich fluids strongly affects the
fireball fluid: It reduces 
its total energy $E_{\scr f}^{\scr{(tot)}}$ as compared to the case
without interaction (i.e. $\tau=$ 100 fm/c). Even the released energy
$E_{\scr f}^{\scr{(released)}}$ drops down. This effect results from
additional stopping of baryon-rich fluids associated with friction
with the fireball fluid. Because of this additional stopping, the
baryon-rich fluids produce less secondary particles (and those
produced are less energetic). Naturally,  
this effect is more pronounced at the
energy 158 A$\cdot$GeV, since the amount of produced secondary
particles is much larger in this case than that at lower energies.

At realistic values of the formation times, $\tau=$ 0.5 and 1 fm/c, the
effect of the interaction is substantially reduced. It happens because the
fireball fluid starts to interact only near the end of the interpenetration
stage. As a result, by the end of the collision
process it looses only 10\% of its available energy
$E_{\scr f}^{\scr{(released)}}$ at $E_{\scr{lab}}=$ 158 A$\cdot$GeV
and 30\%, at $E_{\scr{lab}}=$ 10.5 A$\cdot$GeV. Certainly, this effect
should be observable in mesonic quantities, in particular, in such
fine observables as directed and elliptic flows. The global baryonic
quantities stay practically  unchanged at finite $\tau$. Indeed,
$E_{\scr f}^{\scr{(released)}}$ remains almost the same as at
$\tau=$ 100 fm/c.

The energy content of the baryon-rich fluids exceeds that of the
fireball by an order of magnitude at $E_{\scr{lab}}=$ 158 A$\cdot$GeV
(3440 GeV) and even more at $E_{\scr{lab}}=$ 10.5 A$\cdot$GeV (929
GeV). Therefore, the interaction with the fireball fluid  does
not essentially change the global baryonic quantities. As for refined
baryonic observables,  our preliminary 
calculations of the directed nucleon flow show no changes at
$E_{\scr{lab}}=$ 10.5 A$\cdot$GeV and only slight changes in the
mid-rapidity region at $E_{\scr{lab}}=$ 158 A$\cdot$GeV.
This means that our previous results on the nucleon directed flow and
its excitation function, obtained within the 2-fluid 
model \cite{INNTS}, are not affected by the interaction between the
baryon-rich and  fireball fluids.

\section{Conclusions}

In this paper we have developed a 3-fluid model for simulating
heavy-ion collisions in the  range of incident energies between few to
about 200 A$\cdot$GeV. In addition to two baryon-rich fluids, which
constitute the 2-fluid model
\cite{MRS88,MRS89,MRS91,MRS91a,INNTS}, a delayed evolution of the
produced baryon-free (fireball) fluid is incorporated. This
delay is governed by a formation time, during which
the fireball fluid neither thermalizes nor interacts with the
baryon-rich fluids. After the formation, it thermalizes and comes into
 interaction with the baryon-rich fluids. This interaction is estimated
 from elementary pion-nucleon cross-sections.

The hydrodynamic treatment of heavy-ion collisions is an
alternative to kinetic simulations. The hydrodynamic approach has
certain advantages and disadvantages. Lacking the microscopic
feature of kinetic simulations, it overcomes their basic
assumption, i.e. the assumption of binary collisions, which is
quite unrealistic in dense matter. It directly addresses the
nuclear EoS that is of prime interest in heavy-ion research.
Furthermore, our 3-fluid model uses only friction forces
instead of a vast body of differential cross-sections of
elementary processes, which are generally unknown
experimentally. Naturally, we have to pay for all these pleasant
features of hydrodynamics: the treatment assumes
 that the non-equilibrium stage of the collision can be
described by the 3-fluid approximation. However, all
the assumptions used are quite transparent and can be tested
numerically.

We have simulated relativistic nuclear collisions within  the 3D
code based on the relativistic 3-fluid hydrodynamics combined with
the EoS of the statistical mixed-phase model of the deconfinement
phase transition, developed in~\cite{NST98,TNS98,NTS99}. We
performed calculations of nucleus--nucleus collisions Pb+Pb at
$E_{\scr{lab}}=$ 158 A$\cdot$GeV and Au+Au at $E_{\scr{lab}}=$
10.5 A$\cdot$GeV. To reveal the role of the interaction between
fireball and baryon-rich fluids, we examined the evolution
of global quantities of the fireball fluid.

For zero formation time ($\tau=$ 0) the interaction strongly affects
the fireball fluid: It considerably reduces its  total energy as compared
to that  without interaction. However, for realistic reasonable 
finite formation time, $\tau\approx$ 1 fm/c, the effect of the
interaction is substantially reduced. The fireball fluid
looses only 10\% of its available energy
at $E_{\scr{lab}}=$ 158 A$\cdot$GeV
and 30\%, at $E_{\scr{lab}}=$ 10.5 A$\cdot$GeV. Certainly, this effect
should be observable in mesonic quantities, in particular, in such
sensitive observables like directed and elliptic flows. Since the energy
content of the baryon-rich fluids is much higher than that of the
fireball fluid, global baryonic quantities remain insensitive to this
interaction. As our preliminary calculations show, even the directed
nucleon flow remains practically unaffected by this interaction. In
particular, this fact justifies our previous results on the nucleon
directed flow and its excitation function, obtained within the 2-fluid
model \cite{INNTS}.

\vspace*{5mm} {\bf Acknowledgements} \vspace*{5mm}

We are grateful to L.M. Satarov for useful discussions and careful
reading of the manuscript.
This work was supported in part by the Deutsche Forschungsgemeinschaft
(DFG project 436 RUS 113/558/0-2), the Russian Foundation for Basic
Research (RFBR grant 03-02-04008), Russian Minpromnauki (grant NS-1885.2003.2)
and the German BMBF (contract RUS-01/690).


\begin{thebibliography}{99}

\bibitem{BB56} S.T.~Belyaev and G.I.~Budker,
Dokl. AN SSSR {\bf 107}, 807 (1965).
%
\bibitem{SST02} V.V.~Skokov, S.A.~Smolyansky, and V.D.~Toneev,
 hep-ph/0210099;  V.V.~Skokov, D.V.~Vinnik, S.A.~Smolyansky, and V.D.~Toneev,
  JINR Communications P2-2002-215, Dubna.
%
\bibitem{MRS88}  I.N.~Mishustin, V.N.~Russkikh, and
L.M.~Satarov, Yad. Fiz. {\bf 48}, 711 (1988)
[Sov. J. Nucl. Phys. {\bf 48},  454 (1988)].
.
%
\bibitem{MRS89}  I.N.~Mishustin, V.N.~Russkikh, and
L.M.~Satarov, Nucl. Phys. {\bf A494}, 595 (1989).
%
\bibitem{IMS85} Yu.B.~Ivanov,  I.N.~Mishustin, and L.M.~Satarov,
 Nucl. Phys. {\bf A433}, 713 (1985).
%
\bibitem{IS85}
    Yu.B.~Ivanov and L.M.~Satarov,
     Nucl. Phys. {\bf A446}, 727 (1985).
%
\bibitem{MRS91}   I.N.~Mishustin, V.N.~Russkikh, and
L.M.~Satarov, Yad. Fiz. {\bf 54}, 429 (1991)
[Sov. J. Nucl. Phys. {\bf 54}, 260 (1991)].
%
\bibitem{MRS91a}   I.N.~Mishustin, V.N.~Russkikh, and
L.M.~Satarov, in {\em Relativistic heavy ion physics},
 L.P.~Csernai and  D.D.~Strottman (eds), (World Scientific, 1991) p.179.
%
\bibitem{INNTS}
      Yu.B.~Ivanov, E.G.~Nikonov, W.~N\"orenberg, V.D.~Toneev and
      A.A.~Shanenko,  Heavy Ion Phys. {\bf 15}, 127 (2002)
      [nucl-th/0011004].
%
\bibitem{1-hydro} J.~Sollfrank, P.~Huovinen, M.~Kataja,
  P.V.~Ruuskanen, M.~Prakash and R.~Venugopalan, Phys. Rev. C {\bf  55},
  392 (1997); P.~Huovinen,  P.V.~Ruuskanen and  J.~Sollfrank,
  Nucl. Phys. {\bf A650}, 227 (1999).
%
\bibitem{HS95} C.M.~Hung and  E.V.~Shuryak,  Phys. Rev. Lett. {\bf 75},
4003 (1995);  Phys. Rev. C {\bf 57}, 1891 (1998); D.~Teaney, J.~Lauret,
and E.V.~Shuryak, nucl-th/0110037.
%
\bibitem{Kat93}
U.~Katscher, D.H.~Rischke, J.A.~Maruhn, W.~Greiner, I.N.~Mishustin,
and L.M.~Satarov, Z. Phys. {\bf A346}, 209 (1993);
%
U.~Katscher, J.A.~Maruhn, W.~Greiner, and  I.N.~Mishustin, Z. Phys.
{\bf A346}, 251 (1993);
%
A.~Dumitru, U.~Katscher, J.A.~Maruhn, H.~St\"ocker, W.~Greiner, and
D.H.~Rischke, Phys. Rev. {\bf C51}, 2166 (1995) [hep-ph/9411358];
%
Z. Phys. {\bf A353}, 187 (1995) [hep-ph/9503347].
%
\bibitem{Brac97}
By J.~Brachmann, A.~Dumitru, J.A.~Maruhn, H.~St\"ocker, W.~Greiner, and
D.H.~Rischke, Nucl. Phys. {\bf A619}, 391 (1997) [nucl-th/9703032]; 
%
A.~Dumitru, J.~Brachmann, M.~Bleicher, J.A.~Maruhn, H.~St\"ocker, and
W.~Greiner, Heavy Ion Phys. {\bf 5}, 357 (1997) [nucl-th/9705056];
%
M.~Reiter, A.~Dumitru, J.~Brachmann, J.A.~Maruhn, H.~St\"ocker, and
W.~Greiner, Nucl. Phys. {\bf A643}, 99 (1998) [nucl-th/9806010];
%
M.~Bleicher, M.~Reiter, A.~Dumitru, J.~Brachmann, C.~Spieles,
S.A.~Bass, H.~St\"ocker,  and W.~Greiner, Phys. Rev. {\bf C59}, 1844 (1999)
[hep-ph/9811459];
%
J.~Brachmann, A.~Dumitru, H.~St\"ocker, and W.~Greiner,
  Eur. Phys. J. {\bf A8}, 549 (2000) [nucl-th/9912014];
%
J.~Brachmann, S.~Soff, A.~Dumitru, H.~St\"ocker, J.A.~Maruhn,
W.~Greiner, L.V.~Bravina, and D.H.~Rischke, Phys. Rev. {\bf C61},
024909 (2000) [nucl-th/9908010].
%
\bibitem{SIS200} Conceptual Design Report
{\em `` An International Accelerator Facility for Beams of Ions and
  Antiprotons''}, http://www.gsi.de/GSI-Future/cdr/
%
\bibitem{Lattice} F.~Karsch, Lect. Notes Phys. {\bf 583}, 209 (2002)
[hep-lat/0106019].
%
\bibitem{SPS} C.~Lourengo, Nucl. Phys. A {\bf 698}, 13c (2002). 
%
\bibitem{RHIC} F.~Videbaek (BRAMS Collaboration),
Nucl. Phys. A {\bf 698}, 29c (2002);
W.A.~Zajac (PHENIX Collaboration), ibid., 39c; G.~Roland (PHOBOS
Collaboration), ibid., 54c; J.W.~Harris (STAR Collaboration), ibid., 64c.
%
\bibitem{I87}
Yu.B.~Ivanov,
     Nucl. Phys. {\bf A474},  669 (1987).
%
\bibitem{Sat90} L.M.~Satarov, Yad. Fiz. {\bf 52}, 412 (1990)
[Sov. J. Nucl. Phys. {\bf 52}, 264 (1990)].
%
\bibitem{PPVW93}
Madappa Prakash, Manju Prakash, R.~Venugopalan, and G.~Welke,
Phys. Rep. {\bf 227}, 321 (1993).
%
\bibitem{NST98} E.G.~Nikonov, A.A.~Shanenko, and V.D.~Toneev,
Heavy Ion Phys. {\bf 8}, (1998) 89 [nucl-th/9802018].
%
\bibitem{TNS98}  V.D.~Toneev, E.G.~Nikonov, and A.A.~Shanenko,
in {\em Nuclear Matter in Different Phases and Transitions},
eds. J.-P. Blaizot, X.~Campi, and M.~Ploszajczak (Kluwer Academic
Publishers, 1999) p.309.

\bibitem{NTS99} E.G.~Nikonov, V.D.Toneev, and A.A.~Shanenko,
Yad. Fiz. {\bf 62},  1301
(1999) [Physics of Atomic Nuclei {\bf 62},  1226
(1999)].




\end{thebibliography}
\end{document}